\documentclass[useAMS,usenatbib]{mn2e}
\usepackage{graphicx}

\usepackage{amsmath}
\usepackage{amssymb}

\title[Water and Ice]{Chemistry and Radiative Transfer of Water in Cold, Dense  Clouds}
\author[Keto, Rawlings, Caselli]{Eric Keto$^{1}$,\thanks{E-mail:
keto@cfa.harvard.edu (EK); \hfill\break jcr@star.ucl.ac.uk (JR) \hfill\break
p.caselli@leeds.ac.uk (PC)} Jonathan Rawlings$^{2}$, 
and Paola Caselli$^{3}$
\\
$^{1}$Harvard-Smithsonian Center for Astrophysics, 160 Garden St, Cambridge, MA 02420, USA \\
$^{2}$University College London, London, UK \\
$^{3}$School of Physics and Astronomy, University of Leeds, Leeds LS2 9JT, UK
}
\begin{document}

\date{February 19,  2014}


\maketitle

\label{firstpage}

\begin{abstract}
The Herschel Space Observatory's recent detections 
of water vapor in the cold, dense cloud L1544
allow a direct comparison between observations and
chemical models for oxygen species in 
conditions 
just before star formation. 
We explain a chemical model for gas
phase water, simplified for the limited number of
reactions or processes that are active in extreme cold ($<$ 15 K).
In this model, water 
is removed from the gas phase by freezing onto grains
and by photodissociation. 
Water is formed as ice
on the surface of dust grains from O and OH
and released into the gas phase  by photodesorption. 
The reactions are fast enough with respect to the slow dynamical evolution
of L1544 that  the gas phase water is in equilibrium for the local
conditions thoughout the cloud.
We explain the paradoxical radiative transfer 
of the H$_2$O ($1_{10}-1_{01}$) line.
Despite discouragingly high optical depth
caused by the large Einstein A coefficient, the
subcritical excitation in the cold,
rarefied H$_2$ causes
the line brightness to scale linearly with column
density. Thus the water line can
provide information on
the chemical and dynamical processes in the darkest region 
in the center of a cold, dense cloud.
The inverse P-Cygni profile of the observed
water line generally indicates a contracting cloud. 
This profile is reproduced with a dynamical model 
of slow contraction 
from unstable quasi-static
hydrodynamic equilibrium (an unstable Bonnor-Ebert sphere).

\end{abstract}


\section{Introduction}

Observations of water vapor in the interstellar medium (ISM) by the 
Infrared Space Observatory  \citep{vD1999} and
the Submillimeter Wave Astronomy Satellite (SWAS) \citep{Bergin2000}
show general agreement with chemical models for 
warm ($> 300$ K) conditions in the ISM \citep{Melnick2000,Neufeld2000}.
However, in cold conditions, most of the water is frozen onto dust
grains \citep{Viti2001,vanDishoeck2013}, 
and the production of water occurs mainly on the 
grain surfaces. In order to test chemical models that 
include grain-surface chemistry we used the Heterodyne
Instrument for the Far-Infrared (HIFI) \citep{deGraauw2010} on
the Herschel
Space Observatory to observe the H$_2$O ($1_{10}-1_{01}$) line
in the cold, dense
cloud L1544 \citep{Caselli2010, Caselli2012}.
The first of these two Herschel observations was
made with the wide-band spectrometer (WBS) and
detected water vapor in absorption against the weak 
continuum radiation of dust in the cloud. Follow-up observations with higher spectral
resolution and sensitivity, made with the
high resolution spectrometer (HRS), confirmed the absorption
and detected a blue-shifted emission line that was 
predicted by theoretical modeling \citep{Caselli2010}, but
too narrow to be seen by the WBS in the first observation.

With the better constraints provided by the second observation,
we improved the chemical and
radiative transfer modeling in our previous papers.
We modified the
radiative transfer code MOLLIE to calculate the line emission
in the approximation that the molecule is sub-critically 
excited. This assumes that the collision rate is so
slow that every excitation leads immediately to a radiative
de-excitation and the production of one photon which
escapes the cloud, possibly after many absorptions and
re-emissions, before another excitation. The
emission behaves as if the line were optically thin with
the line brightness proportional to the column density. This
approximation can be correct even at very high
optical depth as long as the excitation rate is slow enough,
C $<$ A/$\tau$, where C is the collision rate, A is the spontaneous
emission rate and $\tau$ the optical depth \citep{Linke1977}.
\citet{Caselli2012} presented the observations and the results
of this modeling.

In this paper, we discuss in detail the theory 
behind the modeling. 
A comparison of the spectral line
observation with theory requires three models.
First, we require
a hydrodynamical model to describe the density, velocity, 
and temperature across
the cloud. We use a model 
of slow contraction in
quasi-static unstable equilibrium 
that we developed in our previous
research 
\citep{KF05,KC10}.
Second, we require
a chemical model to predict the molecular abundance 
across the varying conditions in the cloud.
Following the philosophy for simplified chemical networks in
\citet{KC08} or \citet{BethellBergin2009}, we extract
from a general chemical model for photo-dissociation 
regions \citep{Hollenbach2009} a subset of reactions
expected to be active in cold conditions, principally 
grain-surface reactions as well as freeze-out and 
photodissociation. 
Third, we require 
a radiative transfer model to
generate a simulated molecular line. 
We modify our non-LTE radiative transfer code 
MOLLIE to use the escape
probability approximation. This allows better
control of the solution in extreme optical depth.

The three models are described in more detail in three
sections below. The relevant equations are included in
the appendices.

\section{The three models}

\subsection{The cold, dense clouds}

Given their importance as the nurseries of star formation, the small ($< 0.5$ pc), 
cold ($<15$ K), dense ($n>10^3$ cm$^{-3}$) 
clouds in low-mass 
star ($< 2$ M$_\odot$) forming regions
such as Taurus are widely studied \citep{BerginTafalla2007,diFrancesco2007}. 
Observations show a unique simplicity. 
They contain no internal sources of heat, stars or protostars. 
Their internal turbulence is subsonic, barely broadening
their molecular line widths above thermal \citep{MyersBenson1983}.
With most of their internal energy in 
simple thermal energy, and the weak turbulence 
just a perturbation \citep{Keto2006,Broderick2010}, 
the observed density structure 
approximates the solution of the 
Lane-Emden equation for hydrostatic equilibrium
\citep{Lada2003,Kandori2005}. 
Correspondingly, most are nearly 
spherical with an average aspect ratio 
of about 1.5 \citep{Jijina1999}.  
They are heated from the outside both 
by cosmic rays and by the UV background of starlight
and are cooled from the inside by long wavelength molecular line 
and dust continuum radiation \citep{Evans2001}.
Because of their simplicity, 
we understand the structure and dynamics of these small, cold, dense
clouds
better than any other molecular clouds in the 
interstellar medium. 
They are therefore uniquely 
useful as a laboratory for testing hypotheses 
of more complex phenomena such as the chemistry of 
molecular gas. 

\subsection{Structure and dynamics}

Our physical model for cold, dense clouds is computed with 
a spherical Lagrangian
hydrodynamic code with the gas temperature set by
radiative equilibrium between heating by external
starlight and cosmic rays and cooling by molecular line
and dust radiation. The theory is discussed in \citet{KF05} and
\citet{KC08}. 

\begin{figure}
$
\begin{array}{cc}
\includegraphics[width=3.25in]{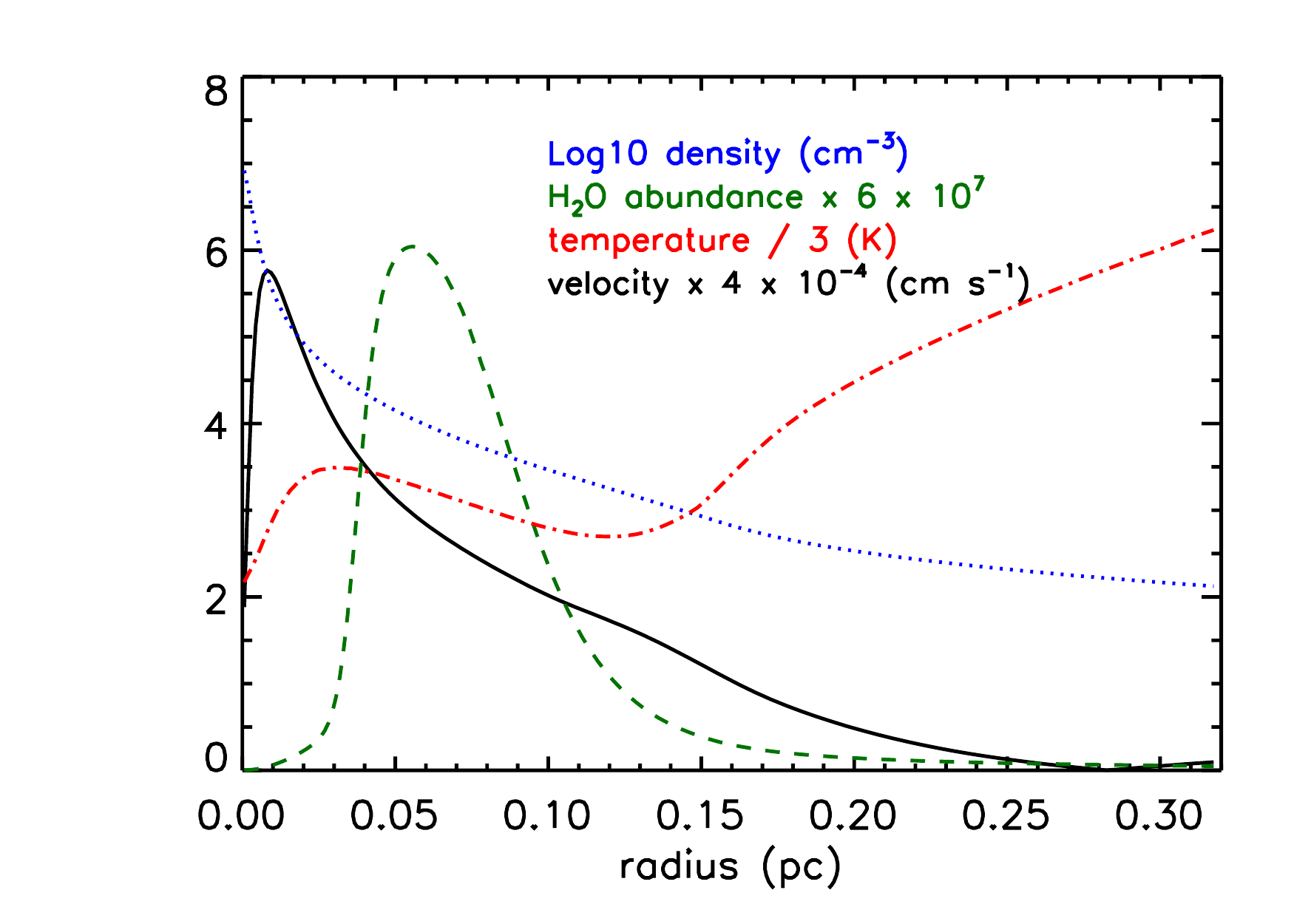}
\end{array}
$
\caption{
Model of a slowly contracting cloud in quasi-static
unstable equilibrium. The log of the density profile in
cm$^{-3}$ is shown in blue (dotted line), the fractional 
abundance of H$_2$O with respect
to H$_2$ is shown in green (dashed line), 
the velocity as the black (solid) line, and
the gas temperature as the red (dot-dashed) line.
The model spectrum is shown in figure \ref{fig:SpectrumLVG}.
}
\label{fig:StructureLVG}
\end{figure}

In our previous research \citep{KC10}, we
generated a dynamical model for the particular case of L1544
by comparing observations and
snapshots in time out of a theoretical model for the 
contraction toward star formation.
We began the hydrodynamic evolution with a 10 M$_\odot$ Bonnor-Ebert (BE) 
sphere with
a central density of $10^4$ cm$^{-3}$ in unstable dynamical equilibrium
and in radiative equilibrium with an external UV field of one Habing flux.
In the early stages of contraction, 
the cloud evolves most rapidly in the center.
As long as the velocities remain subsonic, the evolving
density profile closely follows
a sequence of spherical equilibria or BE spheres with increasing
central densities. 
We compared modeled CO and N$_2$H$^+$ spectra during the 
contraction against those observed in L1544 and
determined that 
the stage of contraction that best matches the data has a central
density of  $1\times 10^7$ cm$^{-3}$and a maximum inward
velocity just about the sound speed \citep{KC10}.
Figure \ref{fig:StructureLVG} shows the density and velocity 
at this time
along with the H$_2$O abundance and temperature.

In the present investigation we modify our numerical hydrodynamic
code to include cooling by atomic oxygen.
This improves the accuracy of the
calculated gas temperature in the photodissociation region
outside the molecular cloud. The equations
governing the cooling by the fine structure lines of atomic oxygen
are presented in the appendix.

\subsection{Chemistry of H$_2$O in cold conditions}

\begin{figure}
\includegraphics[width=3.25in]{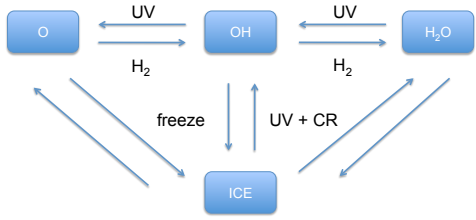}
\caption{Simplified model of the oxygen chemistry in a cold cloud. 
The model includes 3 gas-phase species and H$_2$O ice. 
The significant reactions at cold temperatures (T $<300$ K) are the 
freeze-out of molecules colliding with dust grains, cosmic ray and 
photodesorption of the ice, and photodissociation of the gas phase molecules.}
\label{fig:oxygen_chemistry}
\end{figure}

\begin{figure}
\includegraphics[trim=0.00in 0.0in 0.0in 0.0in, clip, width=3.25in]{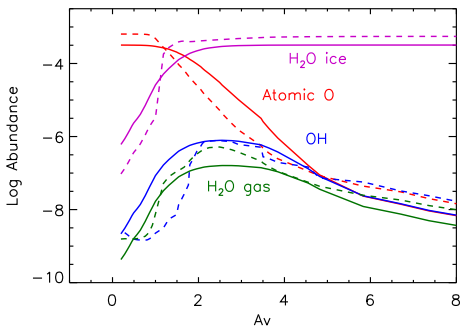}
\caption{Abundances of oxygen species as a function of $A_V$
for the model of L1544 based on a 
slowly contracting Bonnor-Ebert sphere. The figure emphasizes the variation
of abundances in the PDR at the edge.
The figure compares 
the abundances for the physical conditions in (figure \ref{fig:StructureLVG}) from
two models:
\citet{Hollenbach2009} (dashed lines) (courtesy E. Bergin);
and our simplified model (figure \ref{fig:oxygen_chemistry}). 
Figure \ref{fig:oxygenplot-center} shows abundances from the same
models but plotted against log radius to emphasize the variations
in the center.
}
\label{fig:oxygenplot}
\end{figure}

\begin{figure}
\includegraphics[trim=0.00in 0.0in 0.0in 0.0in, clip, width=3.25in]{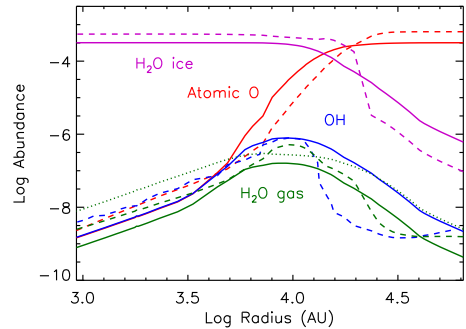}
\caption{Abundances of oxygen species, same as figure \ref{fig:oxygenplot},
except plotted against the log of the radius rather than visual extinction.
This figure emphasizes the variations in abundance in the center.
The figure shows the H$_2$O abundance calculated with our simplified model
using two values for the cosmic ray-induced UV photodesorption 
(equation \ref{eq:CRUV}). The solid green line shows the abundance calculated
with factor $\alpha = 10^{-4}$. The dotted line shows the abundance calculated
with factor $\alpha = 10^{-3}$. The abundance calculated with the 
\citet{Hollenbach2009} model assumes $\alpha = 10^{-3}$ (dashed line).
}
\label{fig:oxygenplot-center}
\end{figure}

The cold conditions in L1544 allow us to simplify the
chemical model for gas phase water. 
We include the four oxygen-bearing species most abundant in cold, dark clouds, 
O, OH, H$_2$O gas, and H$_2$O ice.
Even though all three gas phase molecules may freeze onto the grains, 
we  consider only one species of ice because
the formation
of water from OH and the formation of OH from O are rapid enough on
the grain surface that most of the ice is in the form of H$_2$O. 
To provide a back reaction for the freeze-out of atomic oxygen and
preserve detailed balance, we arbitrarily
assign a desorption rate for atomic O equal to that of H$_2$O even though
the production of atomic O from H$_2$O ice is not indicated. 
Our simplified model 
is shown in 
figure \ref{fig:oxygen_chemistry}. The resulting abundances, calculated
as equilibria between creation and destruction, are shown
in figures \ref{fig:oxygenplot} and \ref{fig:oxygenplot-center}.
Figure \ref{fig:oxygenplot}
shows the abundances near the photodissociation region (PDR) boundary
as a function of the visual extinction, $A_V$. 
Figure \ref{fig:oxygenplot-center} shows the abundances 
against the log of the radius to emphasize the center.

Gas phase water is created by UV photodesorption of water ice which
also creates gas phase OH in a ratio H$_2$O/OH = 2 \citep{Hollenbach2009}. 
In the outer part of the cloud, 
the UV radiation derives from the background field of external starlight. 
The inward attenuation of the UV flux is 
modeled from the visual extinction as $\exp{(-1.8 {\rm A_V})}$. 
In the interior where all the external UV radiation has been attenuated,
the only UV radiation is generated by cosmic ray strikes on H$_2$.
In our previous paper \citep{Caselli2012}, we set this
secondary UV radiation to $1\times 10^{-3}$ times the 
Habing flux (G$_0$=1) \citep{Hollenbach2009}.  In our current model,
we use a lower level, $1\times 10^{-4}$, that is more consistent
with estimated rates \citep{Shen2004}. The difference in abundance for
the two rates is shown in figure \ref{fig:oxygenplot-center}.

H$_2$O and OH are removed from the gas phase by UV photodissociation
and by freezing onto dust grains.
To preserve detailed balance with the photodissociation
we include the back reactions, the gas phase production of H$_2$O,
O + H$_2$ $\rightarrow$ OH and OH + H$_2$ $\rightarrow$ H$_2$O,
even though these are not expected to be important in cold gas.
Removal of gas phase water by freeze-out is 
important in the interior where the higher gas density increases
the dust-gas collision rate, and hence the freeze-out rate.

We assume that 
the gas-phase ion-neutral reactions that lead to the production of water are 
less important 
at cold temperatures ($< 15$ K)
than the reactions that produce water on the
surfaces of ice-coated dust grains.  
Thus, we do not include
gas-phase ion-neutral reactions in the model.
This is valid if the oxygen is quickly removed from the gas-phase by freeze-out
and efficiently converted into water
ice on the grain surface.

By leaving out CO, we avoid coupling in 
the carbon chemistry. Although we already have a simple model for the 
carbon chemistry \citep{KC08,KC10},  we prefer to keep our oxygen model 
as simple as possible. This could create an error of a factor of a few
in the abundance of the oxygen species. Carbon is one-third 
as abundant as oxygen, and in certain conditions CO is the dominant 
carbon molecule. Therefore as much as one-third of
the oxygen could potentially be bound in CO. Ignoring O$_2$ is less of a problem. 
Created primarily by 
the reaction of OH with atomic oxygen, O$_2$ tends to closely follow the 
abundance of OH. Since the amount of oxygen in OH should be 1\% or less
(figure \ref{fig:oxygenplot}),
the abundance of O$_2$ does not affect 
the abundances of the other oxygen species, O, OH, and H$_2$O.

Figures \ref{fig:oxygenplot} and \ref{fig:oxygenplot-center}
compare the abundances from our simplified network with those from
the more complex network of \citet{Hollenbach2009} (courtesy of E. Bergin)
that
includes gas-phase
neutral-neutral and ion-neutral reactions. 
In this calculation, we hold the cloud at the same time in its dynamical
evolution and allow the chemistry to evolve for 10 Myr  
from the assumed starting conditions in which all
species are atomic and neutral. 
Both models generally agree. The gas-phase water
peaks in a region near the
boundary. Here there is enough external UV to rapidly desorb the
water from the ice, but not so much as to dissociate all the molecules.
Further inward, the abundance of water falls as
the gas density and the dust-gas collision
rate (freeze-out rate) both increase while the photodesorption rate decreases with the
attenuation of the UV radiation. 
At high A$\rm _v$, the water is desorbed only by cosmic rays and 
the UV radiation they produce in collisions with H$_2$. 
The general agreement between the two models suggests that the simple model
includes the processes that are significant in the cold environment.
The rate equations for the 
processes selected for the simplified model 
are listed in
the appendix (\S \ref{appendixChemistry}).

\begin{figure}
\includegraphics[trim=0.00in 0.0in 0.0in 0.0in, clip, width=3.25in]{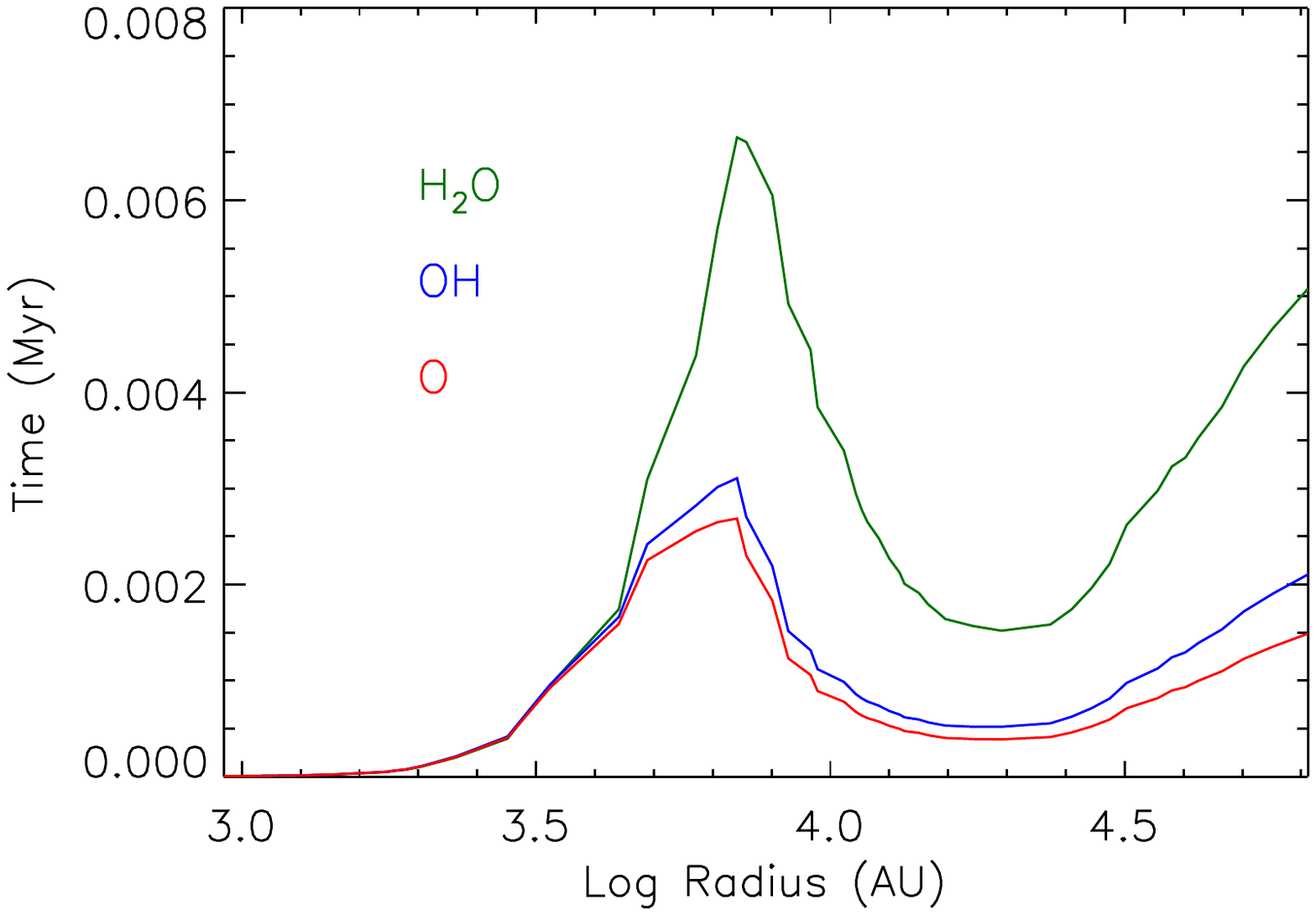}
\caption{
Time scales for chemical equilibrium. From top to bottom, the three lines
show the equilibration time scales for H$_2$O, OH, and O calculated from
equation \ref{eq:time} and the reaction rates in the appendix.
}
\label{fig:time}
\end{figure}

Our simple model calculates equilibrium abundances. We can estimate the
equilibrium time scale from the combined rates for creation and
destruction 
\citep{Caselli2002},
\begin{equation}
t = \frac {t_{creation}t_{destruction}} {t_{creation} + t_{destruction}}
\label{eq:time}
\end{equation}
where the time scales are the inverses of the rates.
Figure \ref{fig:time} shows the equilibrium time scales 
for each species as a function of radius. These may be compared with
the time for the hydrodynamic evolution.
A cloud with a mass of 10 M$_\odot$ and a central
density of $2\times 10^6$ cm$^{-3}$ has a free-fall time,
$t_{ff}=0.03$ Myr
using the central density in the standard equation whereas the sound
crossing time is about 2 Myr \citep{KC10}.
Because the chemical time scales are 
all shorter than the dynamical time scales 
the chemistry reaches equilibrium before the conditions, density, temperature,
and UV flux change.

In this estimate of the time scale for chemical evolution, we are asking
whether the oxygen chemistry in the contracting molecular cloud can maintain 
equilibrium as the cloud evolves dynamically. This is different from the
question of how long it would take for the chemistry to equilibrate
if the gas were held at molecular conditions but evolving from an atomic
state. 

\subsection{Radiative Transfer} \label{RadiativeTransfer}

We use our radiative transfer code MOLLIE \citep{K90,KR10}
to compute model H$_2$O spectra to compare with the Herschel observation.
Here we encounter an interesting question. The large Einstein A
coefficient of the H$_2$O ($1_{10}-1_{01}$) line results in optical depths across
the cloud of several hundred to a thousand depending on excitation. 
High optical depths generally result in
radiative trapping and enhanced excitation of the line. In this case,
the line brightness could have a non-linear relationship to the
column density. For example, the line could be saturated.
On the other hand, the large Einstein A means that the critical
density for collisional de-excitation is quite high ($1\times 10^8$ cm$^{-3}$) 
at the temperatures $< 15$ K, higher than 
the maximum density ($1\times 10^7$ cm$^{-3}$) in our
dynamical model of L1544. This suggests that the line emission
should be proportional to the column density. 

This question was addressed by \citet{Linke1977} who proposed a solution
using the escape probability 
approximation \citep{Kalkofen1984}. 
They assumed a two level molecule, equal statistical
weights in both levels, and the mean radiation field, $\bar{J}$,
set by the escape probability, $\beta$,
\begin{equation} \label{eq:barJ}
\bar{J} = J_0\beta + (1-\beta)S
\end{equation}
where $J_0$ is the continuum from dust and the cosmic microwave
background, $S$ is the line source function, and
\begin{equation} \label{eq:tau}
\beta = (1-\exp{(-\tau)}) / \tau .
\end{equation}
After a satisfying bout with three pages of elementary algebra and
some further minor approximations, they show that as long as $C< A/\tau$,
the line brightness is linearly dependent on the column density, no matter
whether the optical depth is low or high, provided that 
the line is not too bright.

\begin{figure}
$
\begin{array}{cc}
\includegraphics[width=3.25in]{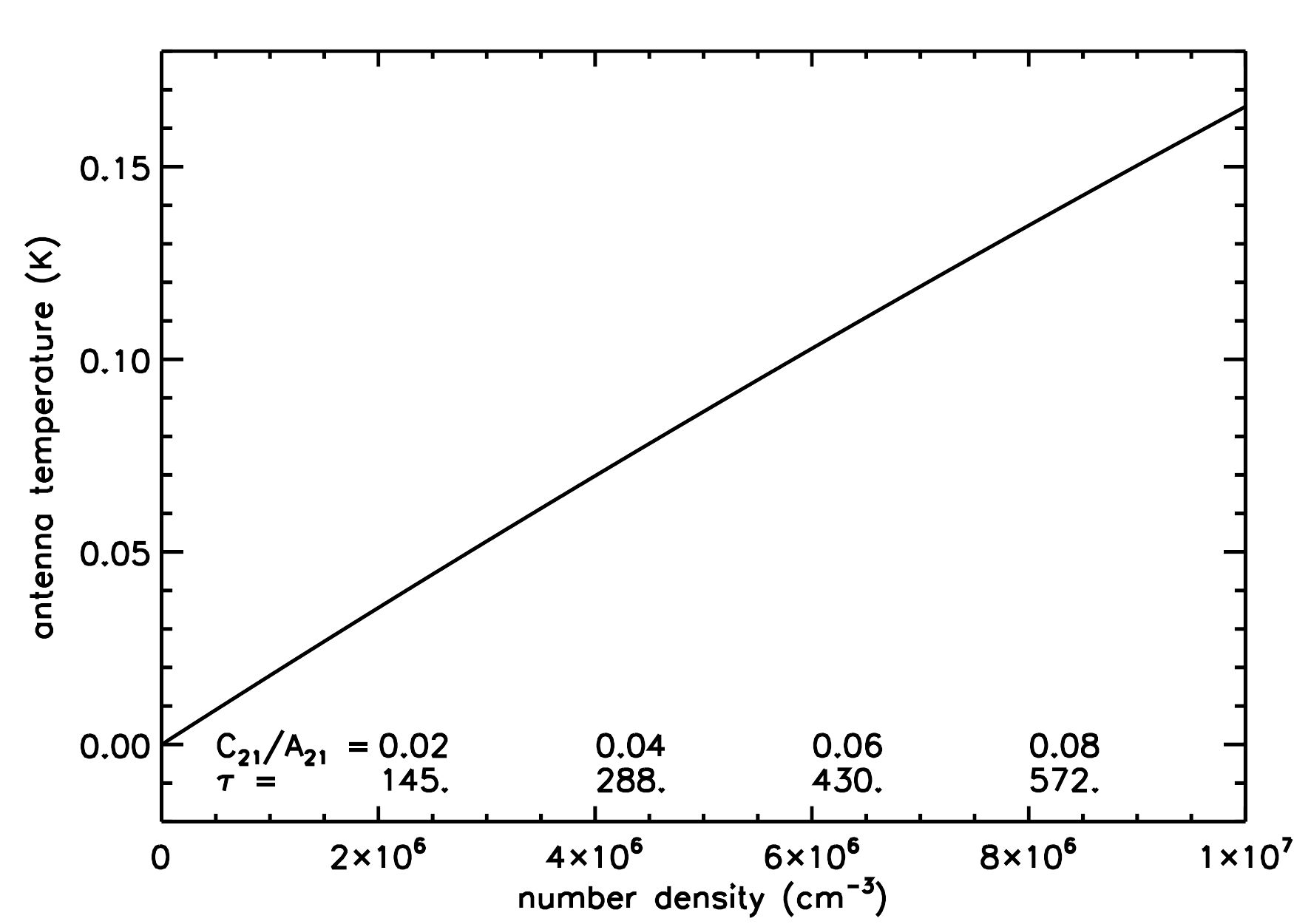}
\end{array}
$
\caption{
The dependence of the observed antenna 
temperature of the H$_2$O ($1_{10} - 1_{01}$) line
on the H$_2$ number density (cm$^{-3}$).
Because the optical depth and the 
ratio of the collision rate to spontaneous emission rate (C/A) are
both linearly dependent on the density, the abscissa can be labeled in
these units as well. Both are shown above the axis. The antenna
temperature is linearly dependent on the density or column density even
at very high optical depth as long as the ratio C/A is small.
}
\label{fig:growth-low3}
\end{figure}

\begin{figure}
$
\begin{array}{cc}
\includegraphics[width=3.25in]{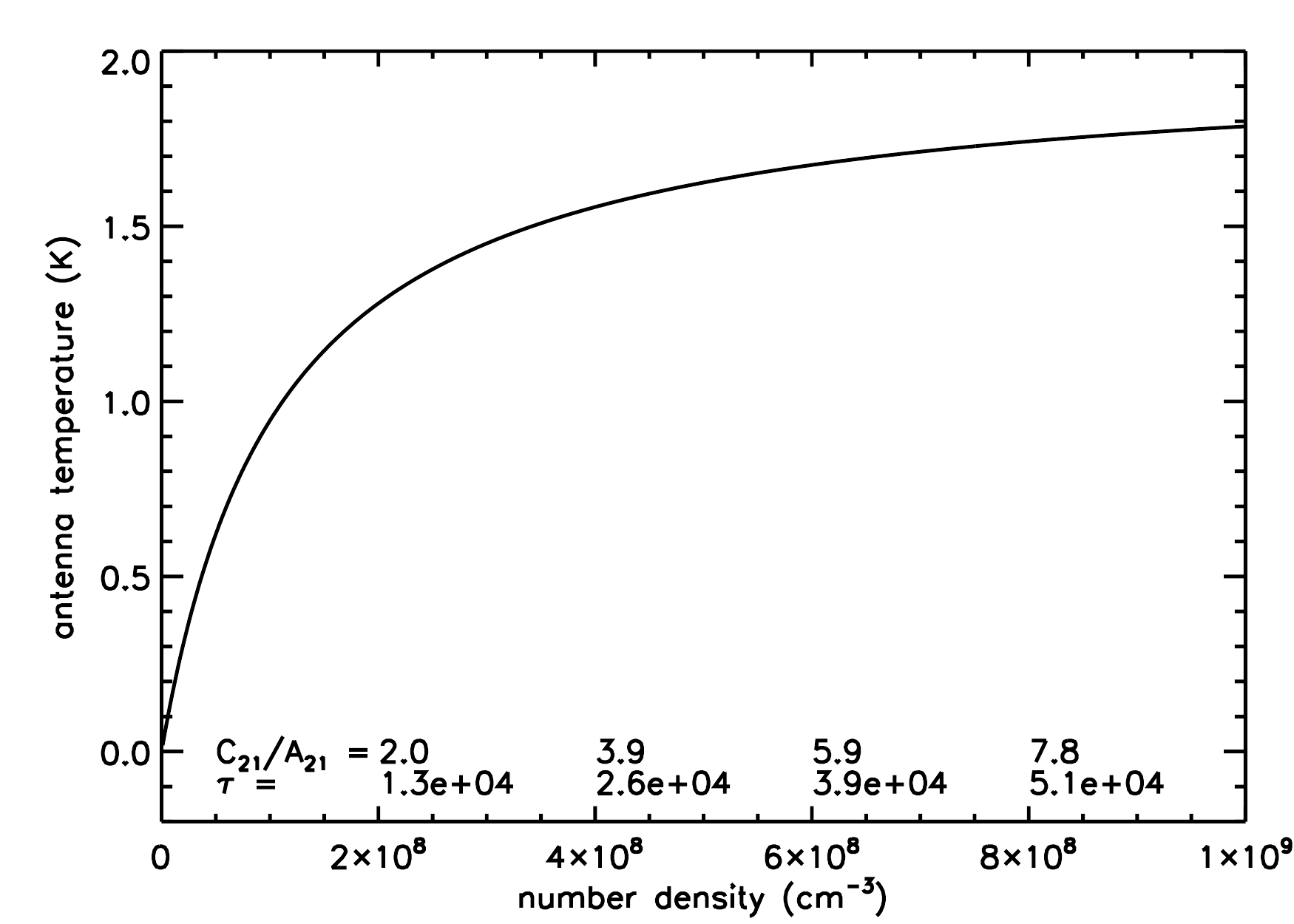}
\end{array}
$
\caption{
The dependence of the observed antenna temperature of the H$_2$O line
($1_{10} - 1_{01}$) on the number density.
Same as figure \ref{fig:growth-low3} but at higher densities where the
ratio C/A is no longer small and the
dependence of the antenna temperature on the density is no longer linear.
}
\label{fig:growth-high3}
\end{figure}

To determine whether the water emission line brightness 
in L1544 has a non-linear
or linear dependence, we numerically 
solve the equations for the two-level molecule with no approximations
other than the escape probability and plot the result. Figures 
\ref{fig:growth-low3} and \ref{fig:growth-high3} show the dependence of the 
antenna temperature on the density for low and high densities respectively. 
Since the column density, the optical depth, and the ratio C/A are 
all linearly dependent on the density, any of these may be used 
on the abcissa. The latter two are shown just above the axis. 
Figure \ref{fig:growth-low3}
shows that the antenna temperature of the water line emission 
is linearly dependent on the
column density even at high density or high optical depth. 
Figure \ref{fig:growth-high3}
shows that the linear relation breaks down when C/A is no longer small.
The densities in both figures show that the water line emission 
in L1544 is in the linear regime.

For an intuitive explanation, suppose that a photon is absorbed on average
once per optical depth of one. A photon may be absorbed and another re-emitted
many times in escaping a cloud of high optical depth. The time scale for 
each de-excitation is $A^{-1}$. Therefore, the time that it takes a photon
to escape the cloud is $\tau /A$. 
As long as this time is shorter than the collisional excitation time ($1/C$), 
then on average, an emitted photon will escape the cloud before another 
photon is created by the next collisional excitation event and 
radiative de-excitation.  In this case, the line remains subcritically excited.
The molecules are in the lower state almost all the time. This is the same 
condition that would prevail if the cloud were optically thin 
($\bar{J}=0$ or $\beta=1$). On this basis, in our earlier paper we
determined the emissivity and opacity of the H$_2$O line
in L1544 by setting $\bar{J}=0$ \citep{Caselli2012}.
This approximation was earlier adopted in analyzing water emission observed by the
SWAS satellite \citep{Snell2000} where it is referred to as "effectively
optically thin".

In this current paper, we seek an improved estimate of 
$\bar{J} > 0$ and $\beta < 1$ by using the escape probability formalism
as suggested by \citet{Linke1977}.
We determine $\beta$ using
the local velocity gradient as given by our 
hydrodynamical model along with the local opacity
using the Sobolev or large velocity gradient (LVG) 
approximation \citep[eqn. 3-40][]{Kalkofen1984}. We use the
6-ray approximation for the angle averaging. 
We allow
for one free scaling parameter on $\beta$ to match the modeled
emission line brightness to the observation. We scale the
LVG opacity by 1/2. 
Because the opacity, column density, 
and line brightness,
are all linearly related, the scaling could be considered
to derive from any or any combination of these parameters.
Given all the uncertain parameters, for example the
mean grain cross-section which also affects the
line brightness (appendix \ref{appendixChemistry}),
this factor of 2 is not significant. 

An alternative method to calculate the excitation is the
accelerated $\Lambda$-iteration algorithm (ALI). 
We do not know if this method is reliable with the
extremely high optical depth, several hundred to a
thousand. 
$\Lambda$-iteration generally converges, but whether it converges to
the correct solution cannot be determined from the algorithm itself
\citep[eqn. 6-33][]{Mihalas1978}.

The excitation may be uncertain, but analysis with the
escape probability method allows us to
understand the effect of the uncertainty.
For example, because we know that the dependence of the line brightness
on the opacity or optical depth is linear, we can say that any
uncertainty in excitation results in the same percentage
uncertainty in the abundance of the chemical model,
or the pathlength of the structural model.

Once $\bar{J}$ is determined everywhere in the cloud,
the equations of statistical equilibrium are solved
to determine the emissivity and opacity.
These are then
used in the radiative transfer equation to
produce the simulated spectral line emission
and absorption. This calculation is done in MOLLIE
in the same way as if $\bar{J}$ were determined
by any other means, for example, by $\Lambda$-iteration.

Both the emissivity and opacity
depend on frequency through the Doppler shifted line
profile function \citep[eqn. 2.14][]{Kalkofen1984}
that varies as a function of position
in the cloud.  We use a line profile function that is the
thermal width plus a microturbulent Gaussian broadening
of 0.08 km s$^{-1}$ derived from
our CO modeling \citep{KC10}. By the approximation of
complete frequency redistribution 
\citep[eqn. 10-39][]{Mihalas1978},
both have the same frequency dependence.
This also implies that each photon emitted
after an absorption event
has no memory of the frequency of the absorbed
photon. It is emitted with the frequency probability
distribution described by the 
line profile function Doppler shifted by the 
local velocity along the direction of emission. 
We also assume complete redistribution in angle.

Figure \ref{fig:SpectrumLVG} shows the modeled line
profile against the observed profile.
The V$_{LSR}$ is assumed to be 7.16 kms$^{-1}$, slightly different
than 7.2 kms$^{-1}$ used in \citet{Caselli2012}. The lower value
is chosen here as the best
fit to the H$_2$O observation.

The combination of blue-shifted emission and
red-shifted absorption is the inverse P-Cygni
profile characteristic of contraction, with
the emission and absorption split by the
inward gas motion in the front and rear of the cloud. 
The absorption against the dust continuum is
unambiguously from the front side indicating
contraction rather than expansion.
This profile has also been seen in other molecules
in other low-mass cold, dense clouds, with the 
absorption against the dust continuum 
\citep{DiFrancesco2001}. 

In L1544, because the inward velocities are below the sound speed,
and the H$_2$O line width is just larger than
thermal,
the emission is shifted with respect to the
absorption by less than a line width. 
In the observations,
what appears
to be a blue-shifted emission line is just the blue 
shoulder and wing of
the complete emission, most of which is brighter,
redder and wider than the observed emission.

Our model also shows weaker emission to the
red of the absorption line. This emission is
from inward moving gas in the front side of
the contracting zone. Again most of the emission
is absorbed by the envelope and only the blue shoulder
of the line is seen. The asymmetry between
the red and blue emission comes about because
the absorbing envelope, which is on the front
side of the cloud, is closer in velocity to
inward flowing gas (red) on the front side of the
contraction. This is the same effect that
produces the blue asymmetric or double-peaked
line profiles seen in contraction 
in molecular lines without such significant envelope
absorption \citep{Anglada1987}. The model shows
more red emission than is seen in the observations.
This red emission may be absorbed by foreground 
gas that is not in the model. Figure 1 of \citet{Caselli2012}
shows additional red shifted absorption in H$_2$O
and red shifted emission in CO, both centered 
around 9 kms$^{-1}$. The blue wing of this red
shifted water line may be absorbing the red
wing of the emission from the dense cloud.

If L1544 were static, no inward contraction,
the emission from the center would be at
the same frequency as the envelope. Because 
of the extremely high optical depth, the absorption
line is saturated and would absorb all the
emission.  We would
see only the absorption line. The depth
of the absorption line is set by brightness
of the dust continuum which is weak (0.011 K)
and not by the optical 
depth of the line which is high
(few hundred to a thousand).

In the current radiative transfer
calculation, we also use a slightly different 
collisional excitation rate 
than before.  The collisional rates for ortho-H$_2$O are 
different with ortho and para-H$_2$.  In our previous 
paper \citep{Caselli2012} we modeled the H$_2$ ortho-to-para 
ratio as a lower limit 1:1 or higher.  Here we assume 
that almost all the hydrogen, 99.9\%, is in the para state. 
This is suggested by recent chemical models that require 
a low ortho-to-para ratio to produce the high deuterium fraction 
observed in cold, dense clouds.
\citep[e.g][]{Kong2013, Sipila2013}.

\section{Interpretation}

\begin{figure}
$
\begin{array}{cc}
\includegraphics[width=3.25in]{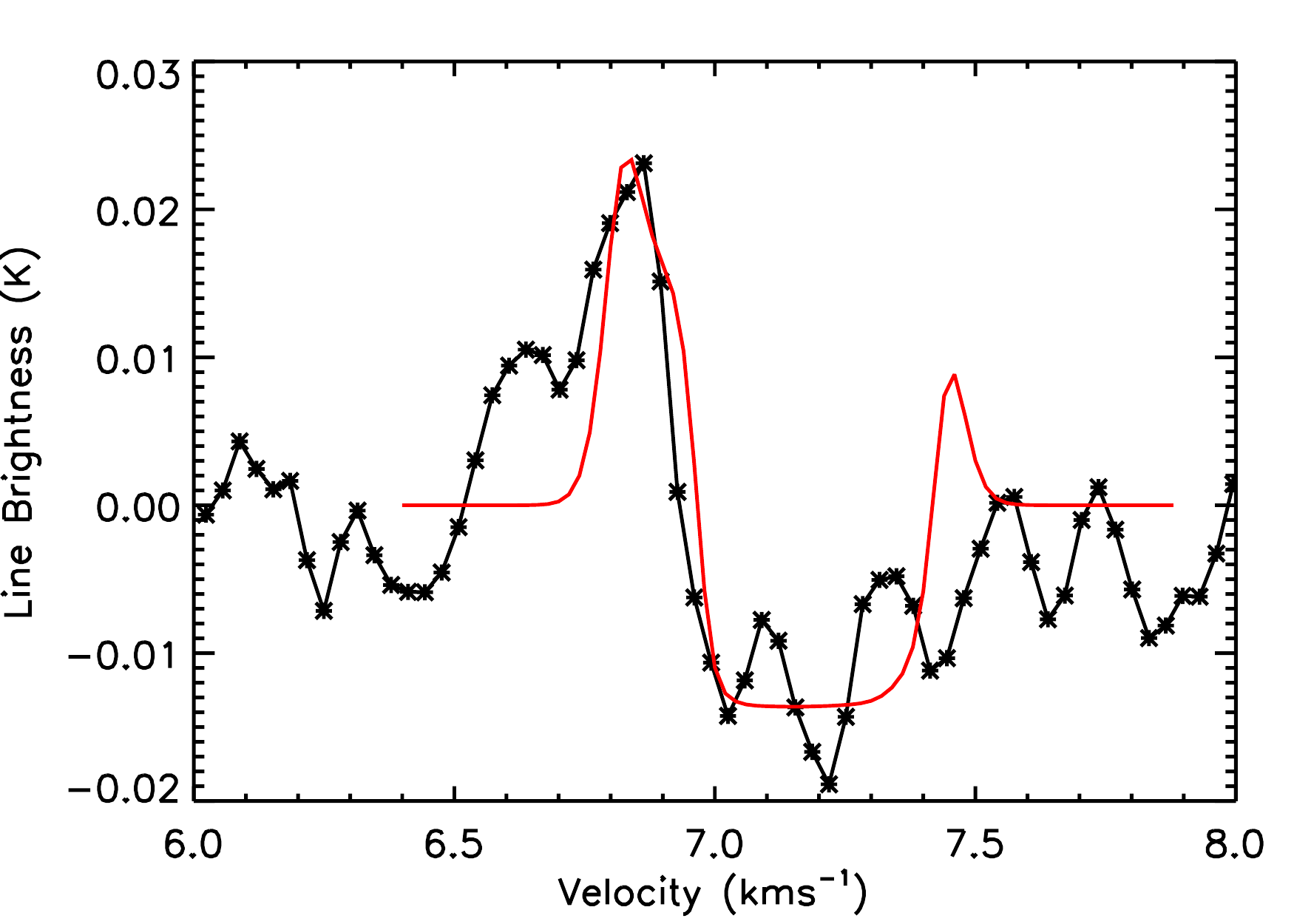}
\end{array}
$
\caption{
Observed spectrum 
of H$_2$O (1$_{10} - 1_{01}$) (black lines with crosses) 
compared with modeled spectrum (simple red line) 
for slow contraction at the time that 
the central density reaches $1 \times 10^7$ cm$^{-3}$.
The model structure is shown in figure \ref{fig:StructureLVG}.
}
\label{fig:SpectrumLVG}
\end{figure}

The shape of the line profile 
(figure \ref{fig:SpectrumLVG})
is unaffected by
any uncertainty in the excitation which scales the emission
across the spectrum. The absorption is saturated and does
not scale with the excitation.
Because of the very
high critical density for collisional de-excitation, 
we know that the line emission is generated only in the densest gas
($>10^6$ cm$^{-3}$)
within a few thousand AU of the center. 
Thus the observation of the inverse P-Cygni profile seen in 
H$_2$O confirms the model for quasi-hydrostatic
contraction with the
highest velocities near the center
(figure \ref{fig:StructureLVG}).

The chemical model requires external UV to create the gas
phase water by photodesorption. This confirms the physical
model of L1544 as a molecular cloud bounded by a photodissociation
region. The UV flux necessarily creates a higher temperature,
up to about 100 K
at the boundary by photoelectric heating. 
This helps maintain the pressure balance at the boundary 
consistent with the model of a BE sphere.

\section{Uncertainties}

The comparison of the simulated and observed spectral line
involves three models each with multiple parameters. Unavoidably 
the choice of parameters in any one of the three models affects 
not only the choice of other parameters in the other two models
but also the interpretation.
It would be a mistake to focus on the uncertainties in any one 
of the models to the exclusion of the others. 
For example, because of the linear relationship between 
the line brightness, the optical depth, and the opacity,
uncertainties in the excitation, pathlength, and abundance,
have equal effect on the spectrum.
A factor of two uncertainty in the excitation can be
compensated by a factor of two in the pathlength or a
factor of two in the abundance of H$_2$O. The pathlength
is unknown.
On the plane of the sky, L1544 has an axial ratio of 2:1,
but we are using a spherical model for the cloud. Our rates
in the chemical model involve estimation of the surface density
of sites for desorption and the covering fraction of water ice
on the grains. The latter is assumed to be one even though we
know that CO and methane ice, not included in the simple model, 
make up
a significant fraction of the ice mantle.
The radiative excitation, parameterized as $\beta$ in
the escape probability is also uncertain because of
the competing effects of high optical depth and subcritical
excitation.

On a linear plot, a factor of two difference in the brightness
of the simulated and observed spectral line looks to be a
damning discrepancy. However, there is at least this much
uncertainty in each of the three models and this does not
significantly affect the conclusions of the study, namely that 
the cloud
can be modeled as a slowly contracting BE sphere bounded
by a photodissociation region with the gas phase water
abundance set by grain surface reactions.

In this paper, we concentrate on the observation of H$_2$O, but 
there are also other constraints that define the model. These
are both observational and theoretical. In
an earlier paper, we showed how observations of CO and N$_2$H$^+$
define the physical model with the two spectral lines giving us
information on the outer and inner regions of the cloud
respectively.
In this regard, the water emission gives us information
in the central few thousand AU 
of the cloud where the density approaches or exceeds the critical
density for de-excitation. This small volume of rapid
inflow and high density does not much affect the N$_2$H$^+$
spectrum which is generated in a much larger volume, and has no 
affect at all on the CO spectrum. A successful model for
L1544 has to satisfy the constraints of all the data.
On the theoretical side, there is an infinite 
space of combinations of abundance, density, velocity, 
and temperature that would form models that match the data. 
Only models that are physically motivated are of interest.
It may be tempting to change the abundances, velocities, or
densities arbitrarily, but this is unlikely to be a 
useful exercise giving the infinite possibilities. A successful
model for L1544 has to be relevant to plausible theory.

There is a natural prejudice for more complex models that
in principle contain more details. 
The goal of our simplified models is to enhance our 
understanding of the most significant
phenomena.
In our research on cold, dense clouds, spanning a number of papers,
we have developed simplified models for the density and temperature
structure, for the dynamics including oscillations, 
for the CO chemistry, and in this paper simplified
models for H$_2$O chemistry and radiative transfer. Each of
these models isolates one or a few key physical processes
and shows how they generate the observables and
operate to control the evolution
toward star formation.

\section{Conclusions}

A simplified chemical model for cold oxygen
chemistry primarily by grain
surface reactions is verified by comparing the simulated spectrum
of the H$_2$O ($1_{10}-1_{01}$) line against an observation of water
vapor in L1544 made with HIFI spectrometer on the Herschel Space
Observatory.

This model reproduces the observed
spectrum of H$_2$O, and also approximates the abundances calculated
by a more complete model that includes gas-phase neutral-neutral and ion-neutral
reactions. 

The gas phase water
is released from ice grains by ultraviolet (UV) photodesorption. 
The UV radiation derives from two sources: 
external starlight and collisions of cosmic rays with
molecular hydrogen. The latter may be important deep inside the
cloud where the visual
extinction is high enough ($>50$ mag) to block out the
external UV radiation.

Water is removed from the gas phase by photodissociation and
freeze-out onto grains. The former is important
at the boundary where the UV from external 
starlight is intense enough to create
a photodissociation region. 
Here, atomic oxygen replaces water as the
most abundant oxygen species.
In the center where the external UV radiation is
completely attenuated, freeze-out is the significant
loss mechanism.

Time dependent chemistry is not required to match the
observations because the time scale for the chemical
processes
is short compared
to the dynamical time scale.

The molecular cloud L1544 is 
bounded by a photodissociation region.

The water emission derives only from the central
few thousand AU 
where the gas density approaches the critical density
for collisional de-excitation of the water line. In the 
model of hydrostatic equilibrium, the gas density in the
center is rising with decreasing radius more steeply than
the abundance of water is decreasing by freeze-out.
Thus the water spectrum provides unique information on the dynamics
in the very center.

The large Einstein A coefficient ($3\times 10^{-3}$ s$^{-1}$)
of the 557 GHz H$_2$O ($1_{10}-1_{01}$) line
results in extremely high optical depth, several hundred to a 
thousand.
However, the density ($< 10^7$ cm$^{-3}$) 
and temperature ($<15$ K) are
low enough that the line is subcritically excited. The 
result is that the line brightness under these conditions
is directly proportional to the column density.

\section{Acknowledgements}

The authors acknowledge Simon Bruderer, Fabien Daniel, Michiel Hogerheijde, 
Joe Mottram, Floris van der Tak for interesting discussions on the radiative transfer of water.
PC acknowledges the financial support of the European Research Council (ERC; project PALs 320620), 
of successive rolling grants awarded by the UK Science and Technology Funding Council.
JR acknowledges the financial support of the Submillimeter Array Telescope.

\bibliography{ms6}

\begin{thebibliography}{42}
\expandafter\ifx\csname natexlab\endcsname\relax\def\natexlab#1{#1}\fi

\bibitem[{{Anglada} {et~al}\mbox{.}(1987){Anglada}, {Rodriguez}, {Canto},
  {Estalella}, \& {Lopez}}]{Anglada1987}
{Anglada} G., {Rodriguez} L.~F., {Canto} J., {Estalella} R., {Lopez} R., 1987,
  \aap, 186, 280

\bibitem[{{Bergin} {et~al}\mbox{.}(2000){Bergin}, {Melnick}, {Stauffer},
  {Ashby}, {Chin}, {Erickson}, {Goldsmith}, {Harwit}, {Howe}, {Kleiner},
  {Koch}, {Neufeld}, {Patten}, {Plume}, {Schieder}, {Snell}, {Tolls}, {Wang},
  {Winnewisser}, \& {Zhang}}]{Bergin2000}
{Bergin} E.~A. {et~al.}, 2000, \apjl, 539, L129

\bibitem[{{Bergin} \& {Tafalla}(2007)}]{BerginTafalla2007}
{Bergin} E.~A., {Tafalla} M., 2007, \araa, 45, 339

\bibitem[{{Bethell} \& {Bergin}(2009)}]{BethellBergin2009}
{Bethell} T., {Bergin} E., 2009, Science, 326, 1675

\bibitem[{{Broderick} \& {Keto}(2010)}]{Broderick2010}
{Broderick} A.~E., {Keto} E., 2010, \apj, 721, 493

\bibitem[{{Caselli} {et~al}\mbox{.}(2012){Caselli}, {Keto}, {Bergin},
  {Tafalla}, {Aikawa}, {Douglas}, {Pagani}, {Y{\'{\i}}ld{\'{\i}}z}, {van der
  Tak}, {Walmsley}, {Codella}, {Nisini}, {Kristensen}, \& {van
  Dishoeck}}]{Caselli2012}
{Caselli} P. {et~al.}, 2012, \apjl, 759, L37

\bibitem[{{Caselli} {et~al}\mbox{.}(2010){Caselli}, {Keto}, {Pagani}, {Aikawa},
  {Y{\i}ld{\i}z}, {van der Tak}, {Tafalla}, {Bergin}, {Nisini}, {Codella}, {van
  Dishoeck}, {Bachiller}, {Baudry}, {Benedettini}, {Benz}, {Bjerkeli}, {Blake},
  {Bontemps}, {Braine}, {Bruderer}, {Cernicharo}, {Daniel}, {di Giorgio},
  {Dominik}, {Doty}, {Encrenaz}, {Fich}, {Fuente}, {Gaier}, {Giannini},
  {Goicoechea}, {de Graauw}, {Helmich}, {Herczeg}, {Herpin}, {Hogerheijde},
  {Jackson}, {Jacq}, {Javadi}, {Johnstone}, {J{\o}rgensen}, {Kester},
  {Kristensen}, {Laauwen}, {Larsson}, {Lis}, {Liseau}, {Luinge}, {Marseille},
  {McCoey}, {Megej}, {Melnick}, {Neufeld}, {Olberg}, {Parise}, {Pearson},
  {Plume}, {Risacher}, {Santiago-Garc{\'{\i}}a}, {Saraceno}, {Shipman},
  {Siegel}, {van Kempen}, {Visser}, {Wampfler}, \& {Wyrowski}}]{Caselli2010}
{Caselli} P. {et~al.}, 2010, \aap, 521, L29

\bibitem[{{Caselli} {et~al}\mbox{.}(2002){Caselli}, {Walmsley}, {Zucconi},
  {Tafalla}, {Dore}, \& {Myers}}]{Caselli2002}
{Caselli} P., {Walmsley} C.~M., {Zucconi} A., {Tafalla} M., {Dore} L., {Myers}
  P.~C., 2002, \apj, 565, 344

\bibitem[{{Conrath} \& {Gierasch}(1984)}]{Conrath1984}
{Conrath} B.~J., {Gierasch} P.~J., 1984, Icarus, 57, 184

\bibitem[{{de Graauw} {et~al}\mbox{.}(2010){de Graauw}, {Helmich}, {Phillips},
  {Stutzki}, {Caux}, {Whyborn}, {Dieleman}, {Roelfsema}, {Aarts}, {Assendorp},
  {Bachiller}, {Baechtold}, {Barcia}, {Beintema}, {Belitsky}, {Benz}, {Bieber},
  {Boogert}, {Borys}, {Bumble}, {Ca{\"i}s}, {Caris}, {Cerulli-Irelli},
  {Chattopadhyay}, {Cherednichenko}, {Ciechanowicz}, {Coeur-Joly}, {Comito},
  {Cros}, {de Jonge}, {de Lange}, {Delforges}, {Delorme}, {den Boggende},
  {Desbat}, {Diez-Gonz{\'a}lez}, {di Giorgio}, {Dubbeldam}, {Edwards},
  {Eggens}, {Erickson}, {Evers}, {Fich}, {Finn}, {Franke}, {Gaier}, {Gal},
  {Gao}, {Gallego}, {Gauffre}, {Gill}, {Glenz}, {Golstein}, {Goulooze},
  {Gunsing}, {G{\"u}sten}, {Hartogh}, {Hatch}, {Higgins}, {Honingh}, {Huisman},
  {Jackson}, {Jacobs}, {Jacobs}, {Jarchow}, {Javadi}, {Jellema}, {Justen},
  {Karpov}, {Kasemann}, {Kawamura}, {Keizer}, {Kester}, {Klapwijk}, {Klein},
  {Kollberg}, {Kooi}, {Kooiman}, {Kopf}, {Krause}, {Krieg}, {Kramer},
  {Kruizenga}, {Kuhn}, {Laauwen}, {Lai}, {Larsson}, {Leduc}, {Leinz}, {Lin},
  {Liseau}, {Liu}, {Loose}, {L{\'o}pez-Fernandez}, {Lord}, {Luinge}, {Marston},
  {Mart{\'{\i}}n-Pintado}, {Maestrini}, {Maiwald}, {McCoey}, {Mehdi}, {Megej},
  {Melchior}, {Meinsma}, {Merkel}, {Michalska}, {Monstein}, {Moratschke},
  {Morris}, {Muller}, {Murphy}, {Naber}, {Natale}, {Nowosielski}, {Nuzzolo},
  {Olberg}, {Olbrich}, {Orfei}, {Orleanski}, {Ossenkopf}, {Peacock}, {Pearson},
  {Peron}, {Phillip-May}, {Piazzo}, {Planesas}, {Rataj}, {Ravera}, {Risacher},
  {Salez}, {Samoska}, {Saraceno}, {Schieder}, {Schlecht}, {Schl{\"o}der},
  {Schm{\"u}lling}, {Schultz}, {Schuster}, {Siebertz}, {Smit}, {Szczerba},
  {Shipman}, {Steinmetz}, {Stern}, {Stokroos}, {Teipen}, {Teyssier}, {Tils},
  {Trappe}, {van Baaren}, {van Leeuwen}, {van de Stadt}, {Visser}, {Wildeman},
  {Wafelbakker}, {Ward}, {Wesselius}, {Wild}, {Wulff}, {Wunsch}, {Tielens},
  {Zaal}, {Zirath}, {Zmuidzinas}, \& {Zwart}}]{deGraauw2010}
{de Graauw} T. {et~al.}, 2010, \aap, 518, L6

\bibitem[{{di Francesco} {et~al}\mbox{.}(2007){di Francesco}, {Evans},
  {Caselli}, {Myers}, {Shirley}, {Aikawa}, \& {Tafalla}}]{diFrancesco2007}
{di Francesco} J., {Evans}, II N.~J., {Caselli} P., {Myers} P.~C., {Shirley}
  Y., {Aikawa} Y., {Tafalla} M., 2007, Protostars and Planets V, 17

\bibitem[{{Di Francesco} {et~al}\mbox{.}(2001){Di Francesco}, {Myers},
  {Wilner}, {Ohashi}, \& {Mardones}}]{DiFrancesco2001}
{Di Francesco} J., {Myers} P.~C., {Wilner} D.~J., {Ohashi} N., {Mardones} D.,
  2001, \apj, 562, 770

\bibitem[{{Evans} {et~al}\mbox{.}(2001){Evans}, {Rawlings}, {Shirley}, \&
  {Mundy}}]{Evans2001}
{Evans}, II N.~J., {Rawlings} J.~M.~C., {Shirley} Y.~L., {Mundy} L.~G., 2001,
  \apj, 557, 193

\bibitem[{{Fouchet}, {Lellouch} \& {Feuchtgruber}(2003){Fouchet}, {Lellouch},
  \& {Feuchtgruber}}]{Fouchet2003}
{Fouchet} T., {Lellouch} E., {Feuchtgruber} H., 2003, Icarus, 161, 127

\bibitem[{{Habing}(1968)}]{Habing1968}
{Habing} H.~J., 1968, Bulletin of the Astronomical Inst. of the Netherlands,
  19, 421

\bibitem[{{Hollenbach} {et~al}\mbox{.}(2009){Hollenbach}, {Kaufman}, {Bergin},
  \& {Melnick}}]{Hollenbach2009}
{Hollenbach} D., {Kaufman} M.~J., {Bergin} E.~A., {Melnick} G.~J., 2009, \apj,
  690, 1497

\bibitem[{{Jijina}, {Myers} \& {Adams}(1999){Jijina}, {Myers}, \&
  {Adams}}]{Jijina1999}
{Jijina} J., {Myers} P.~C., {Adams} F.~C., 1999, \apjs, 125, 161

\bibitem[{{Kalkofen}(1984)}]{Kalkofen1984}
{Kalkofen} W., 1984, {Methods in radiative transfer}

\bibitem[{{Kandori} {et~al}\mbox{.}(2005){Kandori}, {Nakajima}, {Tamura},
  {Tatematsu}, {Aikawa}, {Naoi}, {Sugitani}, {Nakaya}, {Nagayama}, {Nagata},
  {Kurita}, {Kato}, {Nagashima}, \& {Sato}}]{Kandori2005}
{Kandori} R. {et~al.}, 2005, \aj, 130, 2166

\bibitem[{{Keto} {et~al}\mbox{.}(2006){Keto}, {Broderick}, {Lada}, \&
  {Narayan}}]{Keto2006}
{Keto} E., {Broderick} A.~E., {Lada} C.~J., {Narayan} R., 2006, \apj, 652, 1366

\bibitem[{{Keto} \& {Caselli}(2008)}]{KC08}
{Keto} E., {Caselli} P., 2008, \apj, 683, 238

\bibitem[{{Keto} \& {Caselli}(2010)}]{KC10}
{Keto} E., {Caselli} P., 2010, \mnras, 402, 1625

\bibitem[{{Keto} \& {Field}(2005)}]{KF05}
{Keto} E., {Field} G., 2005, \apj, 635, 1151

\bibitem[{{Keto} \& {Rybicki}(2010)}]{KR10}
{Keto} E., {Rybicki} G., 2010, \apj, 716, 1315

\bibitem[{{Keto}(1990)}]{K90}
{Keto} E.~R., 1990, \apj, 355, 190

\bibitem[{{Kong} {et~al}\mbox{.}(2013){Kong}, {Caselli}, {Tan}, \&
  {Wakelam}}]{Kong2013}
{Kong} S., {Caselli} P., {Tan} J.~C., {Wakelam} V., 2013, arxiv:1312.0971

\bibitem[{{Lada} {et~al}\mbox{.}(2003){Lada}, {Bergin}, {Alves}, \&
  {Huard}}]{Lada2003}
{Lada} C.~J., {Bergin} E.~A., {Alves} J.~F., {Huard} T.~L., 2003, \apj, 586,
  286

\bibitem[{{Linke} {et~al}\mbox{.}(1977){Linke}, {Goldsmith}, {Wannier},
  {Wilson}, \& {Penzias}}]{Linke1977}
{Linke} R.~A., {Goldsmith} P.~F., {Wannier} P.~G., {Wilson} R.~W., {Penzias}
  A.~A., 1977, \apj, 214, 50

\bibitem[{{Mathis}, {Rumpl} \& {Nordsieck}(1977){Mathis}, {Rumpl}, \&
  {Nordsieck}}]{MRN1977}
{Mathis} J.~S., {Rumpl} W., {Nordsieck} K.~H., 1977, \apj, 217, 425

\bibitem[{{Melnick} {et~al}\mbox{.}(2000){Melnick}, {Ashby}, {Plume}, {Bergin},
  {Neufeld}, {Chin}, {Erickson}, {Goldsmith}, {Harwit}, {Howe}, {Kleiner},
  {Koch}, {Patten}, {Schieder}, {Snell}, {Stauffer}, {Tolls}, {Wang},
  {Winnewisser}, \& {Zhang}}]{Melnick2000}
{Melnick} G.~J. {et~al.}, 2000, \apjl, 539, L87

\bibitem[{{Mihalas}(1978)}]{Mihalas1978}
{Mihalas} D., 1978, {Stellar atmospheres /2nd edition/}

\bibitem[{{Myers} \& {Benson}(1983)}]{MyersBenson1983}
{Myers} P.~C., {Benson} P.~J., 1983, \apj, 266, 309

\bibitem[{{Neufeld} {et~al}\mbox{.}(2000){Neufeld}, {Snell}, {Ashby}, {Bergin},
  {Chin}, {Erickson}, {Goldsmith}, {Harwit}, {Howe}, {Kleiner}, {Koch},
  {Patten}, {Plume}, {Schieder}, {Stauffer}, {Tolls}, {Wang}, {Winnewisser},
  {Zhang}, \& {Melnick}}]{Neufeld2000}
{Neufeld} D.~A. {et~al.}, 2000, \apjl, 539, L107

\bibitem[{{Rawlings} {et~al}\mbox{.}(1992){Rawlings}, {Hartquist}, {Menten}, \&
  {Williams}}]{Rawlings1992}
{Rawlings} J.~M.~C., {Hartquist} T.~W., {Menten} K.~M., {Williams} D.~A., 1992,
  \mnras, 255, 471

\bibitem[{{Shen} {et~al}\mbox{.}(2004){Shen}, {Greenberg}, {Schutte}, \& {van
  Dishoeck}}]{Shen2004}
{Shen} C.~J., {Greenberg} J.~M., {Schutte} W.~A., {van Dishoeck} E.~F., 2004,
  \aap, 415, 203

\bibitem[{{Sipil{\"a}}, {Caselli} \& {Harju}(2013){Sipil{\"a}}, {Caselli}, \&
  {Harju}}]{Sipila2013}
{Sipil{\"a}} O., {Caselli} P., {Harju} J., 2013, \aap, 554, A92

\bibitem[{{Snell} {et~al}\mbox{.}(2000){Snell}, {Howe}, {Ashby}, {Bergin},
  {Chin}, {Erickson}, {Goldsmith}, {Harwit}, {Kleiner}, {Koch}, {Neufeld},
  {Patten}, {Plume}, {Schieder}, {Stauffer}, {Tolls}, {Wang}, {Winnewisser},
  {Zhang}, \& {Melnick}}]{Snell2000}
{Snell} R.~L. {et~al.}, 2000, \apjl, 539, L101

\bibitem[{{Tielens}(2005)}]{Tielens2005}
{Tielens} A.~G.~G.~M., 2005, {The Physics and Chemistry of the Interstellar
  Medium}

\bibitem[{{Tielens} \& {Hollenbach}(1985)}]{TielensHollenbach1985}
{Tielens} A.~G.~G.~M., {Hollenbach} D., 1985, \apj, 291, 722

\bibitem[{{van Dishoeck} {et~al}\mbox{.}(1999){van Dishoeck}, {Black},
  {Boogert}, {Boonman}, {Ehrenfreund}, {Gerakines}, {de Graauw}, {Helmich},
  {Keane}, {Lahuis}, {Schutte}, {Tielens}, {Whittet}, {Wright}, {van den
  Ancker}, {Blake}, {Creech-Eakman}, {Waters}, \& {Wesselius}}]{vD1999}
{van Dishoeck} E.~F. {et~al.}, 1999, in ESA Special Publication, Vol. 427, The
  Universe as Seen by ISO, {Cox} P., {Kessler} M., eds., p. 437

\bibitem[{{van Dishoeck}, {Herbst} \& {Neufeld}(2013){van Dishoeck}, {Herbst},
  \& {Neufeld}}]{vanDishoeck2013}
{van Dishoeck} E.~F., {Herbst} E., {Neufeld} D.~A., 2013, Chemical Reviews,
  113, 9043

\bibitem[{{Viti} {et~al}\mbox{.}(2001){Viti}, {Roueff}, {Hartquist}, {Pineau
  des For{\^e}ts}, \& {Williams}}]{Viti2001}
{Viti} S., {Roueff} E., {Hartquist} T.~W., {Pineau des For{\^e}ts} G.,
  {Williams} D.~A., 2001, \aap, 370, 557

\end{thebibliography}

\clearpage

\appendix


\section{Cooling by atomic oxygen fine structure lines}

The fine structures lines
of C$^+$ and atomic O
are the major coolants in the diffuse ($n < 1000$ cm$^{-3}$), 
photodissociated gas 
around the molecular clouds. 
The more important coolant at temperatures less than 100 K is C$^+$.
At higher temperatures, oxygen becomes increasingly important 
in the energy balance. The reason is that the 63.2 and 145.6 $\mu$m fine structure lines 
of atomic oxygen have upper states $^3$P$_1$ and $^3$P$_0$  that are at 228 K and 326 K 
above ground, considerably higher than the 92 K of 
the upper state, $^2$P$_{3/2}$ of the 157.6 $\mu$m fine structure line of 
C$^+$.

The cooling by atomic oxygen is simple to model because atomic oxygen 
is 
a product of photodissociation and is therefore abundant only in 
gas with low A$_{\rm v}$ implying gas densities 
below the critical 
densities for collisional de-excitation, 6400 and 3400 cm$^{-3}$ for 
the 63.2 and 145.6 $\mu$m 
lines respectively \citep[table 2.7 of][]{Tielens2005}. At 
this density, we assume that the optically 
thin approximation applies. In this case, every collisional 
excitation to an upper state of the fine structure lines 
results in spontaneous emission that escapes the cloud and cools the gas,
\begin{equation}\label{eq:oxygencooling}
\Lambda_{\rm O} = n({\rm O})n({\rm H_2}) (E_{21}C_{21} + E_{20}C_{20}) 
\ \ {\rm ergs}\ {\rm cm}^{-3}\ {\rm s}^{-1}
\end{equation}
where the upward collision rates are,
\begin{equation}
C_{21} = 1.4\times 10^{-8} 
\frac {g_1} {g_2} C_{12} \exp{(-E_{21} / kT)} \sqrt{T}
\ \ {\rm cm}^3 {\rm s}^{-1} 
\end{equation}
\begin{equation}
C_{20} = 1.4\times 10^{-8}
\frac {g_0} {g_2} C_{02} \exp{(-E_{20} / kT)} \sqrt{T}
\ \ {\rm cm}^3 {\rm s}^{-1}. 
\end{equation}
and the statistical weights are
$g_2 = 5$, 
$g_1 = 3$, and
$g_0 = 1$
and the transition energies are
$E_{12}/k = 228$K
and
$E_{02}/k = 326$K.

\section{Chemistry} \label{appendixChemistry}

\subsection{Freeze-out}
Molecules freeze onto dust grains, sticking when they collide. 
This process is easily modeled. We follow \citet{KC08} 
to calculate the collision timescale.
The time scale for depletion onto dust may be estimated as \citep{Rawlings1992},
\begin{equation}
\tau_{on} = (S_0 R_{dg} n({\rm H_2}) \sigma V_T)^{-1} \ {\rm s}
\end{equation}
Here $S_0$ is the sticking coefficient, with $S_0=1$ meaning that the colliding 
molecule always sticks
to the dust in each collision; $R_{dg}$ is the ratio of the number density of dust grains
relative to molecular hydrogen; $\sigma$ is the mean cross-section of the dust grains;
and $V_T$ is the relative velocity between the grains and the gas.
If the grains have  a power law distribution of sizes with
the number of grains of each size scaling as the -3.5 power of their
radii \citep{MRN1977}, then we can estimate their mean cross-section
as,
\begin{equation}
\langle \sigma \rangle = \bigg(\int^{a_2}_{a_1} n(a)da\bigg)^{-1}\int^{a_2}_{a_1} n(a) \sigma (a) da,
\end{equation}
where $a_1$ and $a_2$ are the minimum and maximum grain sizes. If $a_1 = 0.005$ $\mu$m
and $a_2 = 0.3$ $\mu$m, then $\langle \sigma \rangle = 3.4\times 10^{-4}$ $\mu{\rm m}^2$.
Similarly, the ratio of the number densities of dust and gas may be estimated by
computing the mean mass of a dust grain and assuming the standard gas-to-dust mass
ratio of 100. If the density of the dust is 2 grams cm$^{-3}$, then the ratio of
number densities is $R_{dg} = 4\times 10^{-10}$.

Consistent with \citet{KC08}, our model
has a slightly lower value for the grain cross-section, 
$1.4 \times 10^{-21}$ cm$^2$, than 
\citet{Hollenbach2009},
$\sigma_h = 2 \times 10^{-21}$ cm$^2$. Both values are 
per hydrogen nucleus (2H$_2$ + H). 
Because the ice forms and desorbs off the grain surfaces, 
larger values of the average cross-section result in fewer molecules 
in the gas phase. The actual properties of grains in cold clouds 
are somewhat uncertain.

The relative velocity due to thermal motion is,
\begin{equation}
V_T = \bigg ( {{8kT}\over{\pi\mu}}      \bigg)^{1/2},
\end{equation}
where $T$ is the temperature and $\mu$ the molecular weight.  
The freeze-out rate for species $i$ is,
\begin{equation}
f_i = \tau_{on}^{-1}  n(H_2) \ \ {\rm cm}^{-3}\  {\rm s}^{-1}
\end{equation}

\subsection{Gas-phase reactions}
The neutral-neutral molecular and photodissociation reactions are from 
\citet{TielensHollenbach1985}.
The reaction rate $k_1$ for 
${\rm O + H_2} \rightarrow {\rm OH + H}$ is,
\begin{equation}
k_1 = 3.1 \times 10^{-13} \ (T/300)^{2.7} \exp{(-3150/T)} \ \ {\rm cm}^{3}\ \ {\rm s}^{-1}
\end{equation}
The reaction $k_2$ for ${\rm OH + H_2} \rightarrow {\rm H_2O + H}$ is,
\begin{equation}
k_2 = 2.0 \times 10^{-12} (T/300)^{1.57}\exp{(-1736/T)}  \ \ {\rm cm}^{3}\ \ {\rm s}^{-1}
\end{equation}
 OH + H has a rate,
 \begin{equation}
 5.3 \times 10^{-18} (T/300)^{-5.22} \exp{ (-90/T)  }.
 \end{equation}
All three of these reactions have an activation barrier and are irrelevant at temperatures below 300 K.
The photodissociation rate for the destruction of OH and the formation of O is,
\begin{equation}
P_1 = 3.5 \times 10^{-10} G_0 \exp{(-1.7 A_V)}   \ \ {\rm s}^{-1}
\end{equation}
and the rate for the destruction of H$_2$O and formation of OH is,
\begin{equation}
P_2 = 5.9 \times 10^{-10}  G_0 \exp{(-1.7 A_V)}   \\ \ {\rm s}^{-1}.
\end{equation}
The unitless parameter $G_0 = 1$  corresponds to the 
average local interstellar radiation field in the FUV band \citep{Habing1968}. 
$A_V$ is the visual extinction.

\subsection{Desorption}
The desorption rates are from \citet{Hollenbach2009}. 
The total desorption rate includes thermal desorption, photodesorption,  
and desorption by cosmic rays. We use equation 2 from \citet{Hollenbach2009} for the rate for thermal desorption,
\begin{equation}
D_{Th} = 1.6 \times 10^{11} \bigg( \frac{E_i}{k} \bigg)^{1/2} 
\bigg(\frac{m_H }{ m_i}\bigg)^{1/2} 
\exp{\bigg(\frac{-E_i}{kT_{gr}}\bigg)} \ \ {\rm s}^{-1} \  \ {\rm molecule}^{-1}
\end{equation}
where $E_i/k$, the adsorption energy is 800, 1300, and 
5770 K for O, OH, and H$_2$O respectively, and
m$_i$/m$_{\rm H}$ is the weight of the species with respect to H.  The thermal desorption rate for water
is negligible at the temperatures ($<15$ K) of cold, dense clouds. For the cosmic-ray desorption rate, we use 
equation 8 from \citet{Hollenbach2009}. 
We include only the cosmic-ray desorption rate for H$_2$O, 
\begin{equation}\label{eq:CRdesorption}
D_{CR} =  4.4\times 10^{-17} {\rm molecule}^{-1} {\rm s}^{-1}.
\end{equation}
Both the thermal desorption rate and the cosmic ray desorption rate in units of molecule$^{-1}$ s$^{-1}$ are multiplied
by the number of molecules on the surface of grains per molecule of H$_2$ which is $ N_s  f_s  A_{gr}  R_{dg}   $ where  $N_s = 10^{15} $ cm$^{-2}$ is the number of desorption sites per cm$^2$ on the grain surface
\citep{Hollenbach2009},
$f_s = 1$ is the fraction of the grain surface covered by ice,  the average surface area of a grain is
4 times the grain cross-section, $A_{gr} = 4\sigma = 4 \times 3.4 \times 10^{-4}$ $\mu$m $^2$ \citep{KC08},
and the dust-to-gas ratio $R_{dg} = 4\times 10^{-10}$ \citep{KC08}. 
The photodesorption rates are from equations 6 and 7 \citep{Hollenbach2009},
\begin{equation}\label{eq:UVdesorption}
D_{UV} = G_0  F_0 Y_i f_i \ \exp{(-1.8A_V)}\ {\rm s}^{-1}
\end{equation}
where $F_0 = 10^8$ is the number of UV photons per Habing flux, and 
$Y_i = 10^{-3}$ and $2\times 10^{-3}$ are the 
photodesorption yields per UV photon per second for the production of 
OH and H$_2$O respectively 
from table 1 of \citet{Hollenbach2009}. We  assume 
that all the ice is H$_2$O and follow \citet{Hollenbach2009} in assuming that the photodesorption of this water
ice results in twice as much OH as H$_2$O in the gas phase.

The desorption of water ice does not
result in the production of gas phase oxygen, and
we have no oxygen ice in our model.
To provide a back reaction to the
freeze-out of atomic oxygen, we arbitrarily assign a desorption rate equal to that of water.
In regions of high extinction ($A_V > 4$) this results in a gas phase abundance 
of atomic oxygen that is approximately the same as predicted by \citet{Hollenbach2009}.
This is $< 0.001$ of the total oxygen and has no effect on the other abundances.
In the outer part of the cloud where 
the UV flux is higher ($A_V < 4$) most of the atomic oxygen derives from
photodissociation. Here the UV desorption off grains is insignificant.

Additional desorption is caused by the UV photons emitted by hydrogen
excitation by energetic electrons released in the ionization of hydrogen by
cosmic rays. We follow \citet{Shen2004} and scale this process
as $10^{-4}$ of one Habing flux, $G_0 = 1$, so that,
\begin{equation}\label{eq:CRUV}
D_{CR\ UV} = \alpha G_0  F_0 Y_i f_i \ {s^{-1}}
\end{equation}
with $\alpha = 10^{-4}$.

\subsection{Equilibrium}
In equilibrium, the rate equations in matrix form are,
\begin{equation*}
\begin{array}{lll}
\begin{pmatrix}
-(f_{O} + k_1)	&P_{1}			&0			&0			\\
k_1		&-(f_{OH} + P_1)	&P_{2}			&0			\\
0		&k_2			&-(f_{H_2O} + P_2)	&D_{H_2O}		\\
f_{O} 		&f_{OH}			& f_{H_2O}		&-(D_{OH} + D_{H_2O})	\\
1		&1			&1			&1			\\
\end{pmatrix}
\begin{pmatrix}
O \\
OH \\
H_2O \\
ICE \\
\end{pmatrix} =
\begin{pmatrix}
0 \\
0 \\
0 \\
0 \\
1 \\
\end{pmatrix}
\end{array}
\end{equation*}
where the last row is the conservation equation for oxygen among all the species. As written, this system is overdetermined, but can be solved by dropping any one of the first 4 rows.

\subsection{H$_2$O ortho-para ratio}

Since the ortho state of H$_2$O is 24K above the para state, the O/P ratio in thermal equilibrium is very small at lower temperatures \citep[equation 41][]{Hollenbach2009}.
However, when the water molecule is formed, created from OH on the grain surface for example, it is formed in the ratio of the available quantum states, ortho:para  3:1. The ortho and para states of H$_2$O equilibrate by collisions with H or H$_2$. If the chemical equilibrium time scale is much shorter than the thermal equilibrium time scale, the O/P ratio will not deviate much from 3:1. Observations generally show ratios close to 3:1 \citep{vanDishoeck2013}.

We have not found previous research on the equilibration of H$_2$O, but an appreciation of the
time scale  can be estimated from previous research on the equilibration of the ortho and para states of molecular hydrogen. The dissociation energies of H-H and OH-H are not too different nor the collisional
cross-sections of the molecules. \citet{Conrath1984} and \citet{Fouchet2003} suggest three processes for the equilibration of the ortho and para states of H$_2$ are: (1) gas phase H exchange, (2) gas phase paramagnetic conversion with H$_2$, and (3) H exchange on a surface. We assume that these same processes are applicable to the water. The rates for these processes scale with the gas density through the collision rate and scale as the inverse exponential of the temperature. Scaling the rates for H$_2$ 
from the conditions in the atmosphere of Jupiter to  rarefied, cold gas of the interstellar medium (10 K and $10^6$ cm$^{-3}$) the time scales for these processes are all $> 1$ Gyr.

In contrast, the chemical time scale is very much shorter (figure \ref{fig:time}) throughout the cloud. In this model, water is dissociated in the gas phase by photodissociation and also coming off the grain surfaces by photodesorption in which gas phase OH is produced twice as often as gas phase H$_2$O.
The equilibrium comparison between ortho-para equilibration and chemistry may not be needed because the equilibration time scale exceeds the expected life times of the cold, dense, clouds. 

\end{document}